\documentclass[]{acmart}
\usepackage{pdflscape}
\usepackage{multirow}
\AtBeginDocument{%
  \providecommand\BibTeX{{%
    \normalfont B\kern-0.5em{\scshape i\kern-0.25em b}\kern-0.8em\TeX}}}

\setcopyright{acmcopyright}
\copyrightyear{2023}
\acmYear{2023}

\begin{document}

\title{Multimodal Pathology Image Search Between H\&E Slides and Multiplexed Immunofluorescent Images}

\author{Amir Hajighasemi}
\authornote{Both authors contributed equally to this research.}
\author{Md Jillur Rahman Saurav}
\authornotemark[1]
\affiliation{%
  \institution{The University of Texas at Arlington}
  \country{USA}
}

\author{Mohammad S. Nasr}
\affiliation{%
  \institution{The University of Texas at Arlington}
  \country{USA}
}

\author{Jai Prakash Veerla}
\affiliation{%
  \institution{The University of Texas at Arlington}
  \country{USA}
}

\author{Aarti Darji}
\affiliation{%
 \institution{The University of Texas at Arlington}
 \country{USA}}

\author{Parisa Boodaghi Malidarreh}
\affiliation{%
  \institution{The University of Texas at Arlington}
  \country{USA}
}

\author{Michael Robben, PhD}
\affiliation{%
  \institution{The University of Texas at Arlington}
  \country{USA}
}

\author{Helen H Shang, MD, MS}
\affiliation{%
  \institution{The University of Texas at Arlington;}
  \institution{Ronald Reagan University of California Los Angeles Medical Center}
  \country{USA}
}

\author{Jacob M Luber, PhD}
\affiliation{%
  \institution{The University of Texas at Arlington}
  \country{USA}}
\email{jacob.luber@uta.edu}

\renewcommand{\shortauthors}{Hajighasemi, et al.}

\begin{abstract}
We present an approach for multimodal pathology image search, using dynamic time warping (DTW) on Variational Autoencoder (VAE) latent space that is fed into a ranked choice voting scheme to retrieve multiplexed immunofluorescent imaging (mIF) that is most similar to a query H\&E slide. Through training the VAE and applying DTW, we align and compare mIF and H\&E slides. Our method improves differential diagnosis and therapeutic decisions by integrating morphological H\&E data with immunophenotyping from mIF, providing clinicians a rich perspective of disease states. This facilitates an understanding of the spatial relationships in tissue samples and could revolutionize the diagnostic process, enhancing precision and enabling personalized therapy selection. Our technique demonstrates feasibility using colorectal cancer and healthy tonsil samples. An exhaustive ablation study was conducted on a search engine designed to explore the correlation between multiplexed Immunofluorescence (mIF) and Hematoxylin and Eosin (H\&E) staining, in order to validate its ability to map these distinct modalities into a unified vector space. Despite extreme class imbalance, the system demonstrated robustness and utility by returning similar results across various data features, which suggests potential for future use in multimodal histopathology data analysis.
\end{abstract}




\received{11 June 2023}

\maketitle
\section{Introduction}
Note: A GitHub repository can be accessed with code \href{https://github.com/jacobluber/MultimodalPathologySearch}{here}.

Pathology images, including Hematoxylin and Eosin (H\&E) stained slides, play a crucial role in diagnoses and treatment plans \cite{bhattacharjee2021cluster,kwak2017nuclear,kather2016multi,poojitha2019hybrid,nir2018automatic}. However, the integration of information from multiple imaging modalities, such as immunohistochemistry (IHC) and the recent CODEX technology, can enhance diagnostic precision and treatment effectiveness \cite{Goltsev2018deep}.

Advancements in artificial intelligence and machine learning have enabled efficient analysis and search of multimodal pathology images. Specifically, algorithms have been developed for the extraction of latent space embeddings, revealing the inherent similarities and dissimilarities among different data sources \cite{an2023unicom,tran2021deep}. Despite these advancements, current models still struggle to generalize to diseases they have not been trained on.

Addressing these limitations, we present an approach integrating dynamic time warping (DTW) with Variational Autoencoder (VAE) representations for multimodal pathology image search, aiding in differential diagnosis and therapeutic decisions. We leverage recent contributions in histopathology slide retrieval, like Yottixel \cite{kalra2020yottixel}, SISH \cite{chen2022fast}, and RetCCL \cite{wang2023retccl}, enhancing them with our DTW and VAE approach. We validate our technique using colorectal cancer and healthy tonsil samples and numerous validation experiments.

\begin{figure*}[ht!]
\centering
\includegraphics[width=\textwidth]{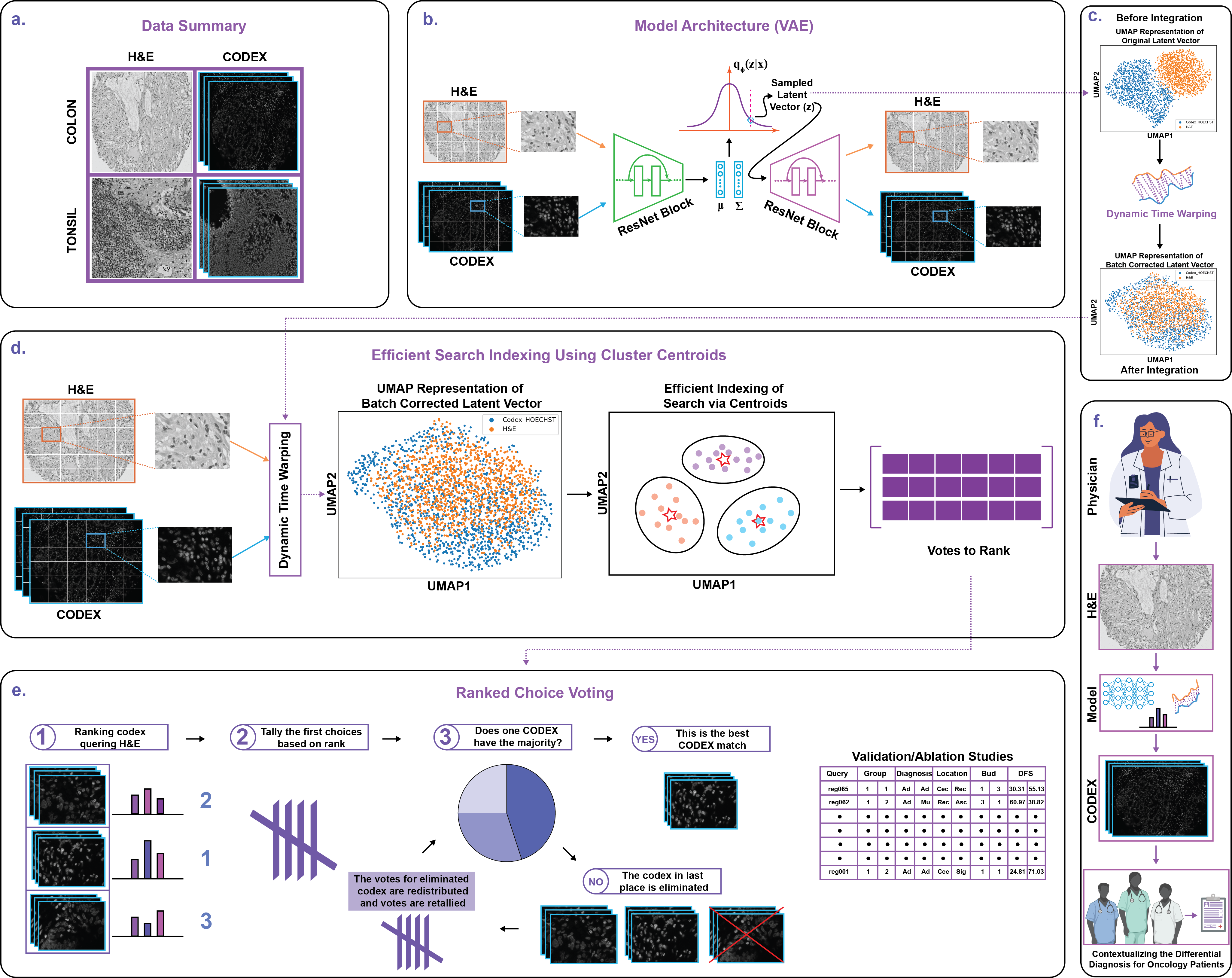}
\caption{ \textbf{1A}: Overview of paired mIF and H\&E data used for training pipeline \textbf{1B}: Multimodal VAE Architecture \textbf{1C} Dynamic Time Warping to integrate latent space of mIF and H\&E patches \textbf{1D}: Using cosine similiarity and centroid based indexing to compare latent spaces accross image modalities \textbf{1E}: Ranked Choice voting to aggregate patch level similiarty to slide level similarity \textbf{1F} Application of our multi-modal search algorithm in the clinic.}
\label{Fig1}
\end{figure*}
Our method presents a novel means of analyzing and comparing tissue samples from different imaging modalities, potentially revolutionizing the diagnostic process. The process is shown in Figure \ref{Fig1}.

\section{Retraining VAE for CODEX Channel Slides}
Building upon previous work in this area \cite{sadegh2023clinically, saurav2022ssim, alsup2023betabuddy, robben2022selection} we successfully retrained the Variational Autoencoder (VAE) to more effectively handle CODEX channel slides. Matched CODEX and H\&E slides from the Human BioMolecular Atlas Program (HubMAP) and previous studies on Tonsil and Colorectal cancer were procured and served as the training dataset for our VAE\cite{black2021codex,Schurch2020coordinated} (Figure 1A and 1B).

The VAE's architecture was adapted to suit the distinct characteristics of CODEX channel slides. Concurrently, we optimized the hyperparameters of the VAE, such as the latent space dimensionality and the reconstruction loss coefficients, to ensure optimal performance. 

Considering the unique characteristics of mIF slides, which are inherently multi-channel data, a simple yet effective approach was employed for channel reduction. Each mIF slide was converted into a series of grayscale images, each representing a specific channel. These grayscale images were then inputted into the VAE one by one. This approach preserves the detailed information in each channel while simplifying the input to the VAE.

Consequently, the training process was tailored to handle the new data form. This adaptation involved tweaking some of the hyperparameters of the VAE, such as the dimensionality of the latent space and the weighting coefficients in the loss function.

The VAE utilizes a weighted loss function consisting of two parts: the reconstruction loss and the Kullback-Leibler (KL) divergence term, as expressed here:
$\mathcal{L} \left(\boldsymbol\theta, \boldsymbol\phi; \boldsymbol{x^{(i)}} \right) = -D_{KL} \left( q_{\boldsymbol\phi} \left( \boldsymbol{z}|\boldsymbol{x^{(i)}} || p_{\boldsymbol\theta} \left( \boldsymbol{z} \right)\right) \right)\  + \mathbb{E}{q{\boldsymbol\phi} \left( \boldsymbol{z}|\boldsymbol{x^{(i)}} \right)} \left[ \log{p_\theta \left( \boldsymbol{x^{(i)}} | \boldsymbol{z} \right)} \right]$. In this equation, $\boldsymbol\theta$ and $\boldsymbol\phi$ are the parameters of the decoder and encoder, respectively, and $\boldsymbol{x^{(i)}}$ is the input slide. The KL term measures the divergence between the encoded distribution $q_{\boldsymbol\phi} \left( \boldsymbol{z}|\boldsymbol{x^{(i)}} \right)$ and a standard Gaussian distribution $p_{\boldsymbol\theta} \left( \boldsymbol{z} \right)$. The second term represents the expected log-likelihood of the data given the latent representation $\boldsymbol{z}$.

In our initial experiments, we used a weighted loss with a KL term coefficient of $0.1$. However, in the revised training process, this coefficient was iteratively tuned depending on the characteristics of the grayscale mIF images. The objective was to find a balance that best maintains the high-dimensional information from mIF while promoting the compact and interpretable latent space that makes the VAE a powerful tool for this task. The latent vector was of dimension 256, batch size was 128, and learning rate was 0.001. 

Post retraining, we generated embeddings of the latent space, represented by the sample latent vector. The dimensionality of the latent space was judiciously chosen to balance compression efficiency with the preservation of essential features. A higher-dimensional latent space yielded superior fidelity in preserving intricate details, while a lower-dimensional space provided a higher compression ratio.

With the latent space embeddings obtained, we conducted cluster analysis to discern the distribution patterns of CODEX and H\&E slides. Visualization of these embeddings revealed distinct clustering of CODEX and H\&E slides, indicating the presence of modality-specific features (Figure 1C).

These observations emphasized the inherent differences between CODEX and H\&E imaging modalities, providing valuable insights. The retraining of the VAE for CODEX channel slides and the subsequent cluster analysis demonstrated the algorithm's ability to capture the unique characteristics of CODEX data while preserving critical information. Thus, validating the suitability of the retrained VAE for subsequent multimodal pathology image search tasks.

\begin{figure}[h]
  \centering
  \includegraphics[width=\linewidth]{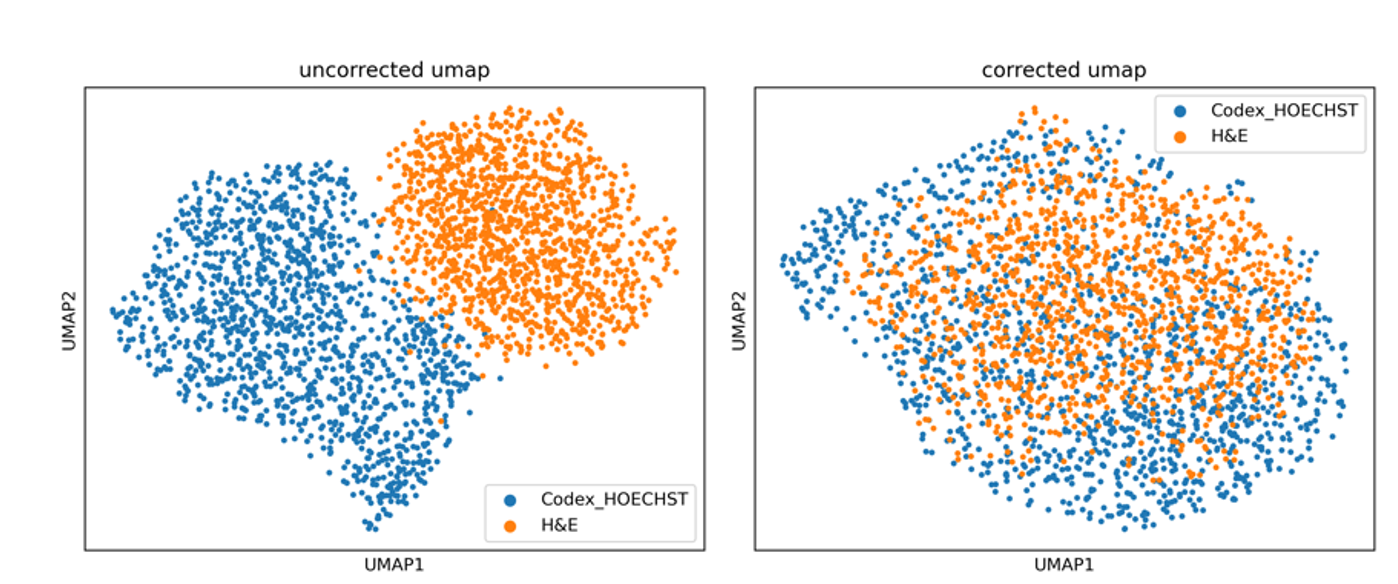}
  \caption{Demonstration of integrating latent space for mIF and H\&E imaging for colorectal cancer. Pre-integration with dynamic time warping is on the left, and post-integration is on the right.}

\end{figure}

\section{Dynamic Time Warping for Matched mIF and H\&E Clustering}

We successfully leveraged the dynamic time warping (DTW) technique to align and cluster matched mIF and H\&E slides represented as latent spaces that have passed through our VAE, yielding a comprehensive analysis of tissue samples. DTW, a robust algorithm that quantifies the similarity between two temporal sequences with varying lengths by warping the time dimension, enabled flexible matching of sequences and accounted for shifts, stretches, and compressions in the temporal domain. 
The objective of DTW is to identify the optimal alignment path that reduces the cumulative distance from the first patch and latent space $(1,1)$ to the final patch and latent space $(n,m)$ in the distance matrix $D$. The optimal path is derived through the following recursive equation:
$$D[i,j] = d(A[i], B[j]) + min(D[i-1, j], D[i, j-1], D[i-1, j-1])$$
where $d(A[i], B[j])$ signifies the distance between $A[i]$ (i-th patch in H\&E sequence) and $B[j]$ (j-th latent space in mIF sequence). This distance metric used was Euclidean distance. The minimum function $\min(D[i-1, j], D[i, j-1], D[i-1, j-1])$ assures the progressive forward and rightward movement of the path, preventing any backward progression.

We applied DTW to the mIF and H\&E slides, aligning their temporal sequences, which facilitated a direct comparison between the two imaging modalities (Figure 1C). This alignment process proved crucial in overcoming the inherent differences in the acquisition and imaging characteristics of mIF and H\&E slides. After alignment, corresponding regions of interest were properly matched, enabling a meaningful analysis (FIgure 1C).

The aligned sequences were then visualized together in the UMAP representation. By representing the aligned mIF and H\&E slides in the UMAP space, we could explore the spatial relationships between the different imaging modalities (Figure 1C and 1D), gaining insights into the similarities and differences in tissue composition.

We developed a search algorithm that could find similar mIF slides to a given input H\&E slide. This algorithm utilized Louvain graph clustering to group mIF slides with similar molecular profiles or morphological features (Figure 1D). For each patch of the inference H\&E slide, we computed cosine similarity to match it with mIF patches and assembled a vector of the five most similar patch mIF images of origin for each patch. This process was repeated for every patch and for every mIF channel, resulting in a 2D array of "votes" that represent the similarity between the H\&E slide and the mIF slides, which we then applied the ranked choice voting (RCV) algorithm to to aggregate patch level similarity into slide level similarity (Figure 1E).

The algorithm returned the most similar multi-channel CODEX slides based on RCV, enabling the identification of mIF slides that were similar to a given H\&E slide. This tool provides a valuable resource for researchers and clinicians to contextualize immune infiltration information captured in H\&E slides but not present in mIF data (Figure 1F).

In summary, the application of DTW alignment and the application of the RCV search algorithm successfully bridged the gap between mIF and H\&E slides, facilitating a comprehensive analysis of tissue samples. The modified voting algorithm improved the accuracy of the search algorithm, enabling the identification of the most similar multi-channel mIF slides to an input H\&E slide.. 

\begin{figure*}
  \centering
  \includegraphics[width=\linewidth]{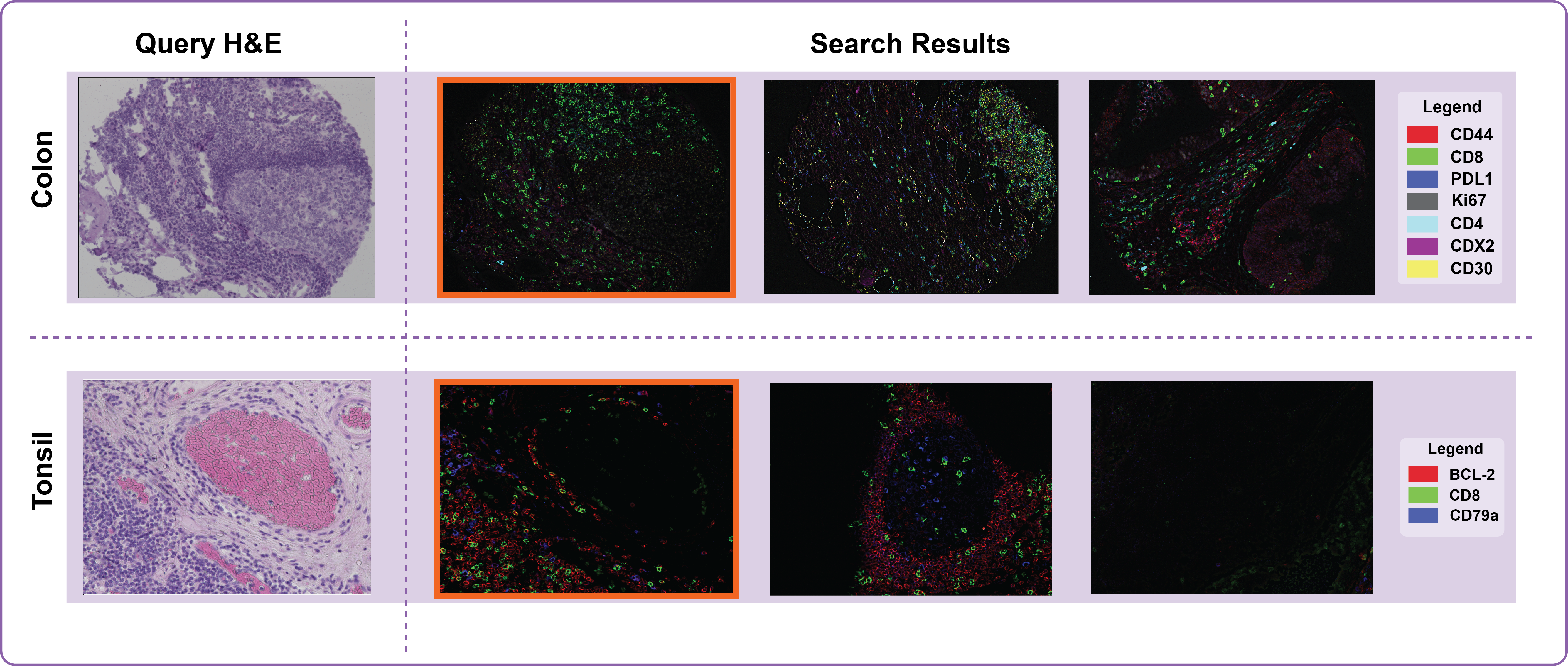}
  \caption{Search Engine Result. A nice validation in addition to our ablation experiments was that for multiple tissues the search engine returns the matching sample for a given H\&E slide (highlighted in orange box) when it is blinded to matching status.}
  \label{fig:your-figure-label}
\end{figure*}

\section{Validation \& Ablation Studies}

To validate our proposed approach to mapping multimodal data, we conducted an exhaustive comparative ablation study using our search engine. The search engine was designed to explore the correlation between two distinct modalities of histopathology: multiplexed Immunofluorescence (mIF) and Hematoxylin and Eosin (H\&E) staining. The aim was to confirm the capability of our system to map these different modalities into a unified vector space, producing meaningful and insightful outcomes.

The primary objective was to establish the efficacy of our methodology by comparing the search outcomes of the two modalities with the five data features represented in our database: Group, Diagnosis, Location, Budding Grade, and DFS. We selected the top two search results for each query and highlighted the correlation between the query and each result. 

Table \ref{tab:validation} presents the results of cross-modal retrieval, showcasing the top 2 search results for mIF images retrieved using H\&E slides as query. The table provides a comparison of clinical information between the query slide and the indexed mIF images. The first subcolumn represents the retrieved first search result, while the second subcolumn represents the clinical information for the second top search result. Noteworthy hits in the clinical information, such as Diagnosis and Budding Group, are highlighted in bold for easy identification.

For the Disease Free Survival (DFS) analysis, instances where the absolute difference between the query and the results fell below 50 were emphasized. This allowed for a precise evaluation of the system's performance, confirming the effectiveness of the multimodal mapping.

The results demonstrate the ability of our system to map different data modalities into a single vector space and extract meaningful information. The comparative study shows that despite the existence of extreme class imbalance, the search engine returned similar results across Group, Diagnosis, Location, Budding Grade, and DFS, each independent from the other. This validation supports the utility and robustness of our system, thereby endorsing its potential for future applications in the field of multimodal histopathology data analysis.

% Validation Table with DFS

\section{Results}

\subsection{Tonsil}
Our approach successfully revealed the complexity and heterogeneity of healthy tonsil tissue. BCL-2, CD8, and CD79a, the markers analyzed in the tonsil tissue, were selected due to their significance in immune response.

BCL-2, a critical regulator of apoptosis, is typically found within the germinal centers of tonsil tissue, representing active regulation of lymphocyte populations. A high expression of BCL-2 suggests inhibition of apoptosis, supporting the maintenance of a robust immune response.

CD8+ cytotoxic T cells play an integral role in immune defense against infections. Analyzing the distribution of CD8+ cells within the tonsil tissue provides insights into the immune responses occurring within this first line of defense against ingested or inhaled pathogens.

CD79a, a marker for B cells, allows the identification of active B-cell populations within the tissue. Understanding the B-cell locations within tonsil tissue can provide insights into adaptive immune responses.

In each case, using mIF to identify similar slides to a given H\&E slide allows for the visualization of these key markers within a spatial context. This approach provides a much more comprehensive understanding of the immune landscape within healthy tonsil tissue than could be achieved with H\&E staining alone (Figure 3).

\subsection{Colorectal Cancer}
Our approach yielded insightful results for colorectal cancer (CRC), a disease known for its heterogeneity. The method visualized the varied expression of key markers such as CD44, CD8, PD-L1, Ki67, CD4, CDX2, and CD30 in the tumor microenvironment.

Simultaneous visualization of CD44, a cancer stem cell marker, and Ki67, a proliferation marker, could indicate regions of highly aggressive tumor growth. CD8+ cytotoxic T cells and CD4+ helper T cells are key players in antitumor immunity. Regions of high CD8 and CD4 expression, especially when coinciding with high PD-L1 expression, could suggest potential responsiveness to immunotherapy, providing valuable guidance for personalized treatment selection. The loss of CDX2, a colorectal differentiation marker, is associated with poor prognosis. By simultaneously observing CDX2 with other markers, the mIF approach can provide a more complete picture of the tumor's pathological state.

By augmenting H\&E slides with mIF data, we provide a richer, more comprehensive view of CRC's complex landscape. This approach adds substantial value in understanding tumor biology, improving diagnostics, and informing personalized treatment strategies for CRC (Figure 3).

\section{Future Work}
The promising results obtained from healthy tonsil tissues and colorectal cancer cases form a strong foundation for the future development of this approach. In our ongoing work, we aim to expand the use of this methodology to more than 20 tissue types. This progression involves a critical step - the computational adjustment of magnification. The aim is to accurately match Hematoxylin and Eosin (H\&E) and multiplexed immunofluorescence (mIF) slides across different scales and resolutions. This flexibility will account for the inherent variabilities seen in tissue preparation and imaging across different laboratories and institutions, thereby enhancing the robustness and broad applicability of our method.

While the theoretical development of this integrative approach holds great value, the practical utility for the medical community is paramount. As such, a key focus of our future work is to develop a user-friendly, intuitive, web-based search engine. This digital platform will serve as an accessible tool for clinicians and medical researchers, allowing them to swiftly and conveniently integrate the H\&E and mIF data. By leveraging our developed search engine, they can gain deeper insights into their diagnostic and therapeutic decision-making process. This form of digital translation is crucial to bringing the benefits of our integrative approach to the frontlines of clinical practice.

\begin{table}
\centering
\begin{tabular}{|p{1cm}|p{0.2cm}|p{0.2cm}|p{0.4cm}|p{0.4cm}|p{0.4cm}|p{0.4cm}|p{0.2cm}|p{0.2cm}|p{0.65cm}|p{0.65cm}|}
\hline
 \multicolumn{1}{|c|}{\textbf{Query}} & \multicolumn{2}{|c|}{\textbf{Group}} & \multicolumn{2}{|c|}{\textbf{Diagnosis}} & \multicolumn{2}{|c|}{\textbf{Location}} & \multicolumn{2}{|c|}{\textbf{Bud}} & \multicolumn{2}{|c|}{\textbf{DFS}} \\
\hline
reg065 & \textbf{1} & \textbf{1} & \textbf{Ad} & \textbf{Ad} & Cec & Rec & \textbf{1} & 3 & \textbf{30.31} & 55.13 \\
\hline
reg062 & 1 & \textbf{2} & Ad & \textbf{Mu} & Rec & \textbf{Asc} & 3 & \textbf{1} & 60.97 & \textbf{38.82} \\
\hline
reg061 & 1 & 1 & \textbf{Mu} & Ad & Cec & Sig & 2 & 3 & 127.67 & 94.80 \\
\hline
reg060 & 1 & 1 & Ad & Ad & \textbf{Sig} & Rec & 1 & \textbf{3} & \textbf{0.19} & \textbf{33.16} \\
\hline
reg057 & \textbf{1} & 2 & \textbf{Ad} & Mu & Rec & Sig & 3 & 3 & \textbf{24.81} & \textbf{30.50} \\
\hline
reg055 & \textbf{1} & \textbf{1} & \textbf{Ad} & \textbf{Ad} & \textbf{Rec} & \textbf{Rec} & \textbf{3} & \textbf{3} & \textbf{22.15} & \textbf{32.97} \\
\hline
reg049 & 1 & \textbf{2} & \textbf{Ad} & \textbf{Ad} & \textbf{Rec} & Cec & 3 & 3 & \textbf{40.18} & \textbf{4.42} \\
\hline
reg046 & 1 & 1 & \textbf{Ad} & Mu & Rec & \textbf{Cec} & \textbf{3} & 2 & \textbf{35.76} & 102.46 \\
\hline
reg036 & 1 & 1 & \textbf{Ad} & Mu & Rec & Cec & \textbf{3} & 2 & 60.78 & 127.48 \\
\hline
reg033 & 2 & 2 & Ad & \textbf{Mu} & Sig & Sig & 3 & 3 & 127.48 & 122.02 \\
\hline
reg029 & \textbf{2} & 1 & \textbf{Ad} & \textbf{Ad} & Rec & Rec & 2 & 3 & \textbf{30.85} & \textbf{48.88} \\
\hline
reg024 & 2 & 2 & Mu & \textbf{Ad} & Asc & Rec & 1 & 2 & 94.80 & 74.01 \\
\hline
reg002 & \textbf{1} & \textbf{1} & \textbf{Ad} & Mu & \textbf{Rec} & Cec & \textbf{3} & 2 & 66.69 & 55.13 \\
\hline
reg001 & \textbf{1} & 2 & \textbf{Ad} & \textbf{Ad} & Cec & Sig & 1 & 1 & \textbf{24.81} & 71.03 \\
\hline
\end{tabular}
\caption{Cross-Modal Retrieval Results of mIF Images from H\&E Slide Query}
\label{tab:validation}
\end{table}

\bibliographystyle{ACM-Reference-Format}
\bibliography{sample-base}

% %%
% %% If your work has an appendix, this is the place to put it.
% \appendix









\end{document}